%% file: Main.tex
\pdfoutput=1 
\documentclass[sigconf]{acmart}
\renewcommand\footnotetextcopyrightpermission[1]{} 
\AtBeginDocument{%
  \providecommand\BibTeX{{%
    \normalfont B\kern-0.5em{\scshape i\kern-0.25em b}\kern-0.8em\TeX}}}

\usepackage{mdframed}
\usepackage{graphicx}
\usepackage{templates/template}
\usepackage{multirow}
\usepackage{graphicx}
\usepackage{enumitem}
\usepackage{wrapfig}

\usepackage{array,multirow,graphicx}
\usepackage{hyperref}

\usepackage{xcolor}
\def\BibTeX{{\rm B\kern-.05em{\sc i\kern-.025em b}\kern-.08em
    T\kern-.1667em\lower.7ex\hbox{E}\kern-.125emX}}
    
\usepackage{amsmath}
\usepackage{booktabs} 
\usepackage{subcaption}
\usepackage{colortbl}
\usepackage{color}

\newcommand{\codefont}[1]{\footnotesize{\texttt{#1}}\normalsize}

\usepackage{listings}
\usepackage{balance}
\usepackage{colortbl}
\usepackage{tcolorbox}
\usepackage{url}
\usepackage{soul}
\usepackage{tcolorbox}

\usepackage[font=small,labelfont=bf]{caption}
\usepackage{tcolorbox}
\definecolor{light-green}{rgb}{.5,1,.5}
\definecolor{light-pink}{rgb}{1,0.5,.5}

\floatstyle{ruled}
\newfloat{example}{thp}{lop}
\floatname{example}{Example }

\newcommand{\hlg}[1]{{\sethlcolor{light-green}\hl{#1}}}
\newcommand{\hlp}[1]{{\sethlcolor{light-pink}\hl{#1}}}

\newcommand{\tool}{UEPerfAnalyzer\xspace}

\usepackage{listings}
\usepackage{setspace}
\begin{document}

\title{Analyzing Performance Issues of Virtual Reality Applications}

\author{Jason Hogan}
\affiliation{%
  \institution{University of Michigan - Dearborn}
  \country{Dearborn, MI, USA}
  }
\email{jnhogan@umich.edu}
  \author{Aaron Salo}
\affiliation{%
  \institution{University of Michigan - Dearborn}
  \country{Dearborn, MI, USA}
  }
  \email{asalo@umich.edu}
\author{Dhia Elhaq Rzig}
\affiliation{%
  \institution{University of Michigan - Dearborn}
  \country{Dearborn, MI, USA}
  }
\email{dhiarzig@umich.edu}

\author{Foyzul Hassan}
\affiliation{%
  \institution{University of Michigan - Dearborn}
    \country{Dearborn, MI, USA}
}
\email{foyzul@umich.edu}

\author{Bruce Maxim}
\affiliation{%
  \institution{University of Michigan - Dearborn}
    \country{Dearborn, MI, USA}
}
\email{bmaxim@umich.edu}



\begin{abstract}
Extended Reality (XR) includes Virtual Reality (VR), Augmented Reality (AR) and Mixed Reality (MR). XR is an emerging technology that simulates a realistic environment for users. XR techniques have provided revolutionary user experiences in various application scenarios (e.g., training, education, product/architecture design, gaming, remote conference/tour, etc.). Due to the high computational cost of rendering real-time animation in limited-resource devices and constant interaction with user activity, XR applications often face performance bottlenecks, and these bottlenecks create a negative impact on the user experience of XR software. Thus, performance optimization plays an essential role in many industry-standard XR applications. Even though identifying performance bottlenecks in traditional software (e.g., desktop applications) is a widely explored topic, those approaches cannot be directly applied within XR software due to the different nature of XR applications. Moreover, XR applications developed in different frameworks such as Unity and Unreal Engine show different performance bottleneck patterns and thus, bottleneck patterns of Unity projects can't be applied for Unreal Engine (UE)-based XR projects.

To fill the knowledge gap for XR performance optimizations of Unreal Engine-based XR projects, we present the first empirical study on performance optimizations from seven UE XR projects, 78 UE XR discussion issues and three sources of UE documentation. Our analysis identified 14 types of performance bugs, including 12 types of bugs related to UE settings issues and two types of CPP source code-related issues. To further assist developers in detecting performance bugs based on the identified bug patterns, we also developed a static analyzer, \tool, that can detect performance bugs in both configuration files and source code. Our evaluation on 37 UE XR projects shows that the static analyzer can detect 428 performance smells in the evaluation projects.

\end{abstract}

%
%



\maketitle
\thispagestyle{empty}

\input{Introduction}
\input{Background}
\input{Approach}
\input{Results}
\input{Implications}
\input{Discussion}
\input{Relatedworks}
\input{Conclusion}

\bibliographystyle{ACM-Reference-Format}
\bibliography{Bibliography}

\end{document}

%% file: Introduction.tex
\section{Introduction}
\label{sec:introduction}
Extended Reality (XR) includes Virtual Reality (VR), Augmented Reality (AR) and Mixed Reality (MR). XR is an emerging technology that simulates realistic environments for users. XR techniques have provided revolutionary user experiences in various application scenarios (e.g., training~\cite{falah2014virtual,chan2010virtual}, education~\cite{serin2020virtual}, product/architecture design~\cite{zwolinski2007use}, gaming~\cite{thomas2012survey}, remote conferencing/tours~\cite{zhang2021xrmas,gunther2018checkmate}, etc.). According to a recent report from MarketsandMarkets Research~\cite{xrmarket}, the XR market size is expected to reach USD 125.2 billion by 2026 from  USD 33.0 billion in 2021, at a Compound Annual Growth Rate (CAGR) of 30.6\% during the forecast period. While the majority of the growth is in hardware areas, the XR software market is also increasing rapidly due to the high demand for software solutions in different domains. In 2019, the AR software market size alone was valued at USD 8.59 Billion~\cite{arsoftwaremarket}. Recent market analysis~\cite{vrsoftwaremarket,mrsoftwaremarket} on VR and MR market research also identified an increasing demand for VR and MR software solutions.

Extended Reality (XR) applications require hardware devices such as smartphone displays, headsets, hand sensors, and software solutions to simulate realistic environments for users. These devices enable users to interact with high-fidelity XR content through hand gestures or hand-tracked controllers. XR software frameworks enable a high-fidelity XR experience through the combination of video displays and tracking devices. Due to the high computational cost of rendering real-time animation on limited-resource devices and constant interaction with user activities, XR applications often face performance bottlenecks. These bottlenecks create negative impacts on user experience with VR/AR software. Some performance bottlenecks may also cause cybersickness and reduced cognitive performance~\cite{mittelstaedt2019vr,stanney2020identifying,stauffert2020latency}. Identifying and fixing performance bugs in XR projects is essential for a  project to succeed.

Identifying performance bottlenecks in traditional software (e.g., desktop applications) is a widely explored topic. These approaches cannot be directly applied to XR projects due to the different nature of XR applications. While XR applications use functional code for rendering, they also use game objects such as graphical animations and audio clips. At the same time, XR applications are often developed using programming frameworks such as the Unity or Unreal game engines. These frameworks often use different programming languages and rendering processes to simulate environments for users. A recent study~\cite{NusratICSE2021} on optimization of VR applications identified common performance bugs and fix patterns for Unity applications. These bug patterns are limited to Unity projects and require manual detection. Like Unity framework, Unreal Engine is a major XR engine powering over 2 million games and boasting a community exceeding 7 million developers. However, there is limited or no existing research on identifying performance bugs in XR applications developed in Unreal Engine (UE).

To fill the knowledge gap of UE XR performance bug patterns, we performed an empirical study on 279 performance optimization changes, 78 issue discussions and three sources of UE documentation. The performance optimization changes are collected from GitHub filtered by XR/UE relevant keywords followed by search using performance relevant keywords and manual verification by the authors. The issue reports documents are collected from UE AnswerHub and documentation by searching performance-related keywords and manual verification by the authors. From this analysis, we identified 14 categories of unique performance-related bug patterns involving source code and configuration files. After that, we developed a novel static analyzer to detect performance bug patterns identified in the prior step. For our tool's evaluation, we applied the static analyzer to 37 UE XR projects and identified new 428 performance bugs in the evaluation projects.

Through our study, we mainly tried to answer the following three research questions. A brief summary of the analysis we identified through the research questions is also presented below.

\begin{itemize}
\item \textbf{RQ1:} What are the major categories of performance bugs in UE-based XR applications?\\
\textbf{Motivation.} The analysis will present a performance bug taxonomy for the UR-based XR applications.\\
\textbf{Result Summary.} Our analysis identified 14 types of performance bugs, among which two performance issues are related to CPP scripts and the rest of the 12 performance issues are related to UE settings.

\item \textbf{RQ2:} What are the major sources of information to identify performance smells?\\
\textbf{Motivation.} The analysis will present the distribution of information sources to identify UE performance smells.\\
\textbf{Result Summary.} Among the identified 25 UE performance smell, 12 smells were identified from OSS commit analysis, two performance smells were identified from UE AnswerHub and ten performance smells were from UE Documentations. This analysis identified that commit analysis and UE Documentation are the most prevalent source of information to know UE XR performance smells.

\item \textbf{RQ2:} How effective is the tool \tool to detect performance smells?\\
\textbf{Motivation.} The analysis will present the effectiveness of the proposed tool \tool.\\
\textbf{Result Summary.} We evaluated \tool on 30 manually labeled smells. Our analysis on the manually labeled smell shows an F-1 score of 0.96. 

\item \textbf{RQ4:} What is the distribution of performance smells detected by \tool?\\
\textbf{Motivation.} The analysis will present the distribution of smells detected by \tool on 37 UE-based XR evaluation projects.\\
\textbf{Result Summary.} Our analysis on 37 evaluation projects shows that \tool detected 49 CPP script-related smells and 379 settings smells. Overall each of the projects contains more than 11 performance smells.

\end{itemize}

Overall, our work presented in the paper makes the following contributions.

\begin{itemize}
    \item A dataset of performance optimization changes, issue discussions and documentation related UE XR projects for future research.
    
    \item Presented an UE XR performance bug taxonomy than can be beneficial for developers and researchers.
    
    \item A novel approach and tool, \tool, for the automatic detection of performance bugs in UE XR projects.
\end{itemize}

The remainder of this paper is organized as follows. After discussing the background in Section~\ref{sec:background}, we discuss the details of the approach in Section~\ref{sec:approach}. Section~\ref{sec:results} presents an evaluation of the approach, while Section~\ref{sec:implications}, Section~\ref{sec:threatsvalidity} and Section~\ref{sec:relatedworks} discuss the implications of the work, the threats to validity and the related works respectively. Finally, we conclude our work with Section~\ref{sec:conclusion}.  
\\

%% file: Background.tex
\section{Background}
\label{sec:background}
\par{This section focuses on giving the reader a brief description of key background subjects that are pertinent to this study.}
\subsection{Extended Reality (XR)}
Extended reality (XR) is an umbrella term used to describe emerging technologies that combine the physical and virtual worlds in three ways to interact with an environment:  Virtual Reality (VR), Augmented Reality (AR), and Mixed Reality (MR).  Each of these forms relies on unique hardware, such as mobile devices, cameras, and stereoscopic headsets, to present simulated experiences to users. Moreover, individual devices have their own set of unique restrictions that must be balanced to produce an optimal performance. An example of this for VR is stereoscopic headsets, which need to render the real-time graphical interface two separate times. Each eye must be shown a slightly different view of the virtual environment that can be changed with the user’s real-time input while still maintaining high frame rates. This ensures users remain immersed in the experience and avoid negative effects from experience (nausea, headaches, etc.)~\cite{XRproblems}. Maintaining this balance comes with high computational costs that need to be considered during development. Other input devices (e.g., motion controllers) contribute to a unique development process where consideration of efficiency is key to maintaining the high performance needed for VR technology. Augmented reality (AR) is an interactive experience taking place in a real-world environment where physical objects are enhanced by computer-generated real-time information, sometimes across multiple sensory modalities, including visual, auditory, etc. Mixed reality (MR) allows us to interact with elements of both the real objects and the digital objects. 

\subsection{Unreal Engine}
Unreal Engine (UE) is one of the most widely used graphics engines in the industry, especially for the development of high-end XR and 3D games. Part of the reason for this popularity is that it is open-source. This allows developers to customize the engine itself. The UE engine code is written in C++, which provides highly portable features for the development of games and XR applications. Currently, the UE development platform supports Windows, MacOS, Linux and FreeBSD, and applications created in UE could be deployed in Microsoft Windows, PlayStation, Xbox One, macOS, iOS, Android, XR, Linux, HTML5, etc. When looking at the engine from an XR standpoint, it has a fully-featured XR Mode inside the editor that allows the user to develop without leaving the XR environment~\cite{UEXR}. A UE project holds all the content, such as Blueprints, Objects, Actors, Controllers, etc., for XR applications. UE provides two toolsets: C++ script support and Blueprint visual scripting, which can be used in tandem to define XR applications workflow. UE allows developers to define display properties with various configuration settings.

\subsubsection{XR Workflow} In UE, user interface elements (e.g. gameplay class, Slate, Canvas) can be implemented as C++ scripts. The changes made through C++ scripts will be reflected in the UE editor after code compilation with Visual Studio or XCode. Apart from defining workflow with C++ scripts, developers can utilize the Blueprint visual scripting system that allows developers to create an entire application without writing a single line of code. Moreover, Blueprint-specific markup is available in UE's C++ implementation, allowing developers to use C++ code and Blueprint~\cite{UEBlueprint} programs in tandem. Blueprints are created visually in the UE editor and stored as assets in content packages. However, UE Blueprint asset files are stored in UE-specific file format and don't provide support for external tools to read and analyze Blueprint assets. In this study, we could not analyze Blueprint asset-related performance issues. 

\subsubsection{Settings for UE Display and Engine Behavior}
Unreal Engine provides support for defining display and engine properties through a set of settings (.ini) files. The DefaultEngine.ini and DefaultGame.ini files are created whenever an XR application is created. UE provides utilities through which script classes can read the configuration files and can display their contents. Listing~\ref{list:exampleconfig} shows such an example where the \codefont{PhysicsSettings} script class loads display properties from the settings file. Like this example, other workflow scripts can load and display configurations from the settings files. Since UE settings (.ini) files are widely used for defining display settings, many performance issues are related to properties defined in these files. In our analysis, we considered the performance issues originating from the configuration files.

{\fontsize{11}{11}
\begin{lstlisting}[columns=flexible,basicstyle=\small,caption=DefaultEngine.ini Configuration Segment from \textit{appodeal/appodeal-ue4-demo}, frame=tblr, captionpos=b, label={list:exampleconfig},escapechar=!]
...
[/Script/Engine.PhysicsSettings]
DefaultGravityZ=-980.000000
DefaultTerminalVelocity=4000.000000
DefaultFluidFriction=0.300000
SimulateScratchMemorySize=262144
RagdollAggregateThreshold=4
TriangleMeshTriangleMinAreaThreshold=5.000000
bEnableAsyncScene=False
bEnableShapeSharing=False
...
\end{lstlisting}
}


%% file: Approach.tex
\section{Approach}
\label{sec:approach}
In this section, we first discussed the dataset and then discussed the performance smell analysis approach, including commit analysis, issue analysis and documentation analysis. Finally, we discussed the automatic performance bug detection tool \tool.

\begin{figure*}[!htbp]
\centering
 \includegraphics[width=0.9\linewidth,height=5.7cm ]{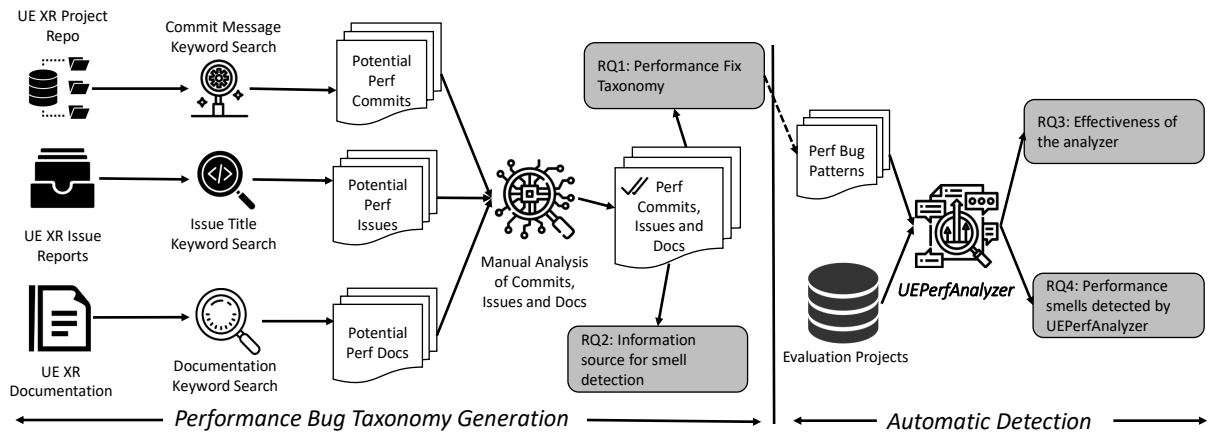}
 \caption{Overview of Research Methodology}
 \label{fig:res_overview}
\end{figure*}

\subsection{Dataset}
To collect the data-set of XR projects, we followed the method proposed by Gonzalez et al.~\cite{gonzalez2020} and adapted it to the context of our work. First, we collected the labels related to \textit{Unreal ar}, \textit{Unreal mr}, \textit{Unreal vr} and \textit{Unreal xr}. Second, we curated the topic labels by removing those which were related to a more generic group of projects than Unreal-Engine-Powered Extended-Reality software projects, such as \textit{Open-GL}. Third, we collected the list of projects labeled with the topic labels collected in the previous step; then we curated this list of projects by removing any which were not related to our specific context. We applied the curation process on the previously collected set 176 projects. After the curation, we identified 45 projects that are related to XR. Finally, we applied commit message search technique proposed by Nusrat et al.~\cite{NusratICSE2021}  utilizing Java Git client JGit to identify possible performance bug-fix commits. The keywords used for performance commit fixing were: performance, speed up, accelerate, fast, slow, latency, contention, optimize, efficient and fps. After applying the filtering process described above, we identified seven projects with 279 commits that contained performance-related keywords. The rest of the 38 UE XR projects were used for evaluation purposes. Moreover, we collected 78 UE XR discussion issues from UE AnswerHub~\cite{UEAnswerHub} and three sources of UE documentation using keyword-based search approach similar to GitHub commit searching.   

\subsection{Identify Performance Fixes, Issues and Documentation}
To identify potential performance related code smells in an XR UE project a set of commit messages, questions posted on the UE4 AnswerHub~\cite{UEAnswerHub}, and the UE documentation were analyzed. The set of criteria that determines if a commit message, a question post, or a documentation piece is a performance issue is:

\begin{enumerate}
    \item Does the issue being addressed relate to VR/AR/XR UE4 projects specifically?
    \item Is the issue related to the performance of the applications?
    \item Is there a clearly identified solution to this issue?
\end{enumerate}

\subsubsection{Commit Analysis}
By analyzing the commit messages of XR UE projects on GitHub, we can identify code smell fixes that the developers performed related to performance issues. We looked for commit messages that were related specifically to performance fixes that affected runtime performance of the game. We had identified a total of 7 projects related specifically to XR UE games and obtained a total of 279 commits that contained performance related keywords in commit messages through an automated process. We then manually analyzed these commit messages to validate if they were in-fact performance related, and we narrowed the list down to 73 commit messages that met our criteria. The 208 commit messages that were not deemed to be valid either contained one or more of the commit message search-keywords but did not contain changes reflecting performance fixes or the changes were not related to the UE VR/AR/XR elements of the project.

Of the 73 commit messages that were deemed to be related to performance fixes through manual analysis, we identified 14 commits that contained possible code smells related to performance issues. Out of these 14 commits, we identified six categories of performance issues that could be detected using the proposed static analysis tool \tool. These include four setting issues, and two CPP code smells.

\subsubsection{Issue Analysis}
The UE AnswerHub~\cite{UEAnswerHub} provides a place for developers to post questions and issues that can be answered by engine developers and community members. The person who asked the question can also mark the answer that solved their problem as the best answer so that other developers can benefit from the question post. By analyzing posts on the UE AnswerHub, we can identify potential code smells that users are encountering and determine if they are related to XR performance issues. The steps we took in order to identify question posts related to XR performance code smells were:

\begin{enumerate}
    \item Define a search query that yielded a sufficient number of results to manually sift through:
    \begin{enumerate}
        \item The first search query utilized “[tags]” and a search string: “Performance [VR] [AR] [XR] [c++]” This search was done on all categories of the UE4 AnswerHub. Unfortunately, it yielded several hundred results that did not specifically pertain to XR applications. When the same search query was attempted within the VR category of the AnswerHub, it yielded only three results.
        \item The second search query broadened the search within the VR category and only contained the following query: “Performance”. This yielded a total of 78 results that seemed very promising. These were the search results that we used to expand the dataset.
    \end{enumerate}
    \item Once a set of search results was defined, each result was manually assessed for its validity and relationship to VR/AR/XR performance code smells. Only posts that had an "Accepted Answer" were considered in an effort to filter out posts that did not contain quality answers. Many of the results proposed questions that only applied to the developer's specific projects and wouldn't necessarily provide benefit to XR projects as a whole. There were also some others which did not necessarily pertain to performance issues but just contained the “performance” keyword. Therefore, these types of posts were filtered out.
\end{enumerate}

After manually assessing the posts and narrowing them down to 12 from the initial 78, we then categorized them into categories similar to those used for assessing the commit messages. Of these issues, we identified one code smell category that could be detected through static analysis. In our analysis, we didn't consider categories related to binary UAsset files, since as those files that are not human-readable and can't be parsed for analysis by any tools other than UE.

\subsubsection{Documentation Analysis}
To further enhance that dataset, we applied a keyword search approach to UE VR/AR/XR-related documentation from well-known sources of UE docs. These sources included “docs.UnrealEngine.com”, “docs.Microsoft.com”, and “developer.oculus.com”. These results were then manually analyzed and cross-referenced against each other to determine what can be seen as possible code smells. We identified A total of 12 possible code smell categories, 8 of them were found to be detectable through static analysis. They consisted of 1 CPP code smell and seven setting smells.

\subsubsection{Analysis Summary}
Combining the identified code smells from the commit analysis, issue analysis, and documentation analysis, a total of 13 code smell categories were identified. Of these 13 smells, two are CPP code smells and 11 are settings-related smells. Identifying performance smells is a complicated task and requires UE and XR application development knowledge. To overcome this challenge; three co-authors manually analyzed identified commits, issues and documents to identify performance smells. Among the three co-authors, two authors are experienced in UE-based XR application development and one author has experience in CPP-based application development. To avoid bias, the authors analyzed separately and then did three rounds of consensus meetings to resolve their disagreements. Finally, Cohen’s Kappa coefficient was calculated, and the coefficient value is 1.00, which depicts a perfect agreement between co-authors.

\subsection{Automatic Performance Bug Detection}
\par{To perform automatic detection of these identified code smells, a static analysis utility was developed. The utility consists of three core components: Settings file analyzer, CPP static analysis, and a main analysis client that ties everything together. The proof of concept for this utility was developed using C++ as a command-line executable application that will analyze a given project directory. For this proof of concept, we implemented 11 settings smell checks and 2 CPP code smell checks.}

\subsubsection{Settings File Analysis}
\par{UE settings files are mostly written as INI-style configuration files and to analyze these INI files, we utilize an open-source project called simpleini~\cite{simpleini} that allows us to easily read the key and value pair of specific sections in the files. The INI analysis was developed in a dynamic method that allows for easy integration and modification of INI checks. The INI checks will analyze all files with an ".ini" file extension in the project directory to determine that proper settings exist with the correct values to allow for smooth XR runtime. See the sample below of an INI check definition:}

\begin{lstlisting}
INICodeSmell* iniSmell_renderer_settings
    = new INICodeSmell("/Script/Engine.RendererSettings",
                       "vr.InstancedStereo",
                       "True",
                       "DefaultEngine.ini",
                       "Instance Stereo",
                       "InstancedStereo should be set to true in DefaultEngine.ini",
                       false);
\end{lstlisting}

\subsubsection{CPP Analysis}
\par{To analyze CPP files in the specified project directory for code smells, we utilize the srcML~\cite{srcml} library to generate an abstract syntax tree of each CPP file. This allowed us to easily traverse the AST of the CPP file to detect if a specified code smell exists in it. To detect the existence of a code smell, we identified key context points within the code that are indicative of the code smells exist. For example, when checking for the "Motion Controller Code Smell" we the following context details to determine its existence:}
\begin{itemize}
    \item Was the "CreateDefaultSubobject" method called?
    \item Was the object created of the type "UMotionControllerComponent"?
    \item Was "bDisableLowLatency" set to "True" on this object after its creation?
    \begin{itemize}
        \item If so the code smell does not exist
        \item If not, the code smell does exist
    \end{itemize}
\end{itemize}
These checks are executed on every CPP file detected in the project directory.

\subsubsection{Analysis Client}
\par{The analysis client component of this utility is what ties the INI and code smell checks together. When executing the utility, a project's directory is passed to it as a launch parameter and is then be searched for all .ini and .cpp files. First, all the .ini files are checked using the methods described above. Then all .cpp files are analyzed for the code smells we defined. Issues detected within .ini or .cpp files are recorded with the name of the issue, its location, and a suggestion to resolve it. After execution of the analysis is completed, a "results.json" file is saved to the project directory containing all of the identified issues. By savomg the results in a neatly formatted .json file, our tool allows developers to generate performance smell in a convenient manner to allow detailed analysis and fixes. Below is a sample of the generated results for an .ini and a .cpp issue:}

\begin{lstlisting}
{
    "ProjectPath" : "/home/projects/UE4VR/",
    "TotalCppFilesChecked" : 7,
    "TotalIniFilesChecked" : 3,
    "FoundCodeSmells" : [
        {
            "FilePath" : "/home/projects/UE4VR/Config/DefaultEngine.ini",
            "CodeSmellName" : "Instance Stereo",
            "CodeSmellSuggestion" : "InstancedStereo should be set to true in DefaultEngine.ini",
            "LineNumber" : 0
        },
        {
            "FilePath" : "/home/projects/UE4VR/Source/VRCharacter.cpp",
            "CodeSmellName" : "Motion Controller Code Smell",
            "CodeSmellSuggestion" : "Not setting bDisableLowLatencyUpdate to false can cause performance issues.",
            "LineNumber" : 0
        }
    ]
}
\end{lstlisting}

%% file: Results.tex
\section{Results}
\label{sec:results}
\subsection{RQ1: Categorization of XR Performance Optimizations}
\label{subsec:rootcause}

\begin{figure*}[!htbp]
\centering
 \includegraphics[width=0.9\linewidth ]{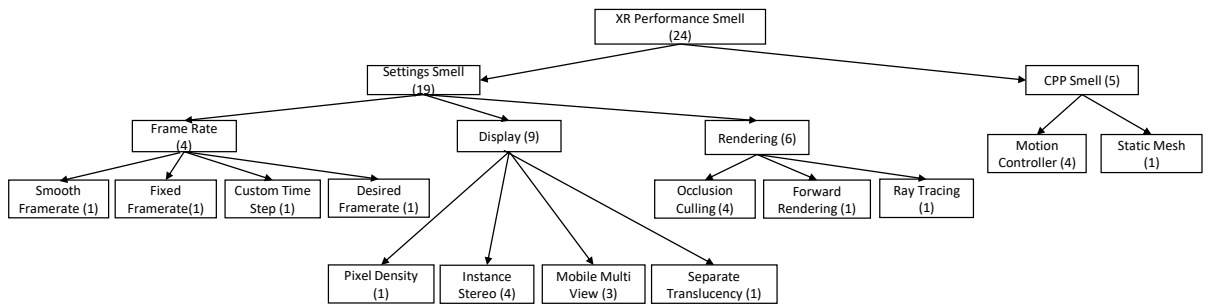}
 \caption{XR Performance Optimization Taxonomy}
 \label{fig:taxonomy}
\end{figure*}

In total, our research identified 13 distinct categories of performance-related code smells that can be detected through static analysis. They were derived from our commit analysis, issue analysis, and documentation analysis. The full taxonomy is shown in figure~\ref{fig:taxonomy}. Settings smells and CPP smells have their own subcategories which further identify the code smells associated with these concepts.

\subsubsection{Settings} 
Code smells in this category are related to Unreal Engine’s project settings and can be resolved through the project’s .ini files. Our findings show that settings-related smells carry the biggest weight when it comes to performance in XR applications developed with Unreal Engine. Settings-related smells can be separated into three subcategories: Framerate, Display, and Rendering. Each of these categories is discussed as follows:\newline

\noindent\textbf{Frame Rate (4 smells)}
Frame rate-related code smells effect the way the end-user perceives the environment. Low frame-rates and/or stuttering can cause a loss of immersion or motion sickness~\cite{motionsickness}. Most XR headsets have a 90hz refresh rate, so ensuring an application runs at 90+ fps is ideal~\cite{desiredframerate}. The following setting adjustments should be verified when considering framerate-related code smells: disabling \textbf{Smooth framerate}, disabling \textbf{Fixed Framerate}, disabling \textbf{Custom Timestep}, and adjusting \textbf{Desired Framerate}. Listing~\ref{list:framerate} shows such as Framerate related fix where a developer updated Min Desired Framerate to improve XR application performance while maintaining XR application quality.


{\fontsize{11}{11}
\begin{lstlisting}[columns=flexible,basicstyle=\small,caption=An Example of Frame Rate Fix \footnotesize{(davidjcc/HorrorVR, 4384ede)}, frame=tblr, captionpos=b, label={list:framerate},escapechar=!]
...
!\hlp{- SmoothedFrameRateRange=(LowerBound=(Type=Inclusive,Value=60.000000), UpperBound=(Type=Exclusive,Value=120.000000))}!
!\hlg{+ SmoothedFrameRateRange=(LowerBound=(Type=Inclusive,Value=120.000000), UpperBound=(Type=Exclusive,Value=120.000000))}!
...
!\hlp{- MinDesiredFrameRate=60.000000}!
!\hlg{+ MinDesiredFrameRate=120.000000}!
...
\end{lstlisting}
}

\noindent\textbf{Display (4 smells)}
Display-related code smells can affect the performance of the unique displays used by XR applications, including VR headsets and mobile devices. We found four major smells related to this. This first is the adjustment of the \textbf{Pixel Density} to 1 as this considered the ideal setting when working with modern VR headsets. Anything less than or greater than this will either lower the sharpness of the elements visible to the user or lower the performance of the application~\cite{motionsickness}. The second and third are to enable \textbf{Instanced Stereo} and \textbf{Mobile Multiview} which enables rendering the VR headset image for both eyes in parallel. This is less taxing than rendering each eye separately and increases the application's performance~\cite{desiredframerate}. The final display smell is disabling \textbf{Separate Translucency}. This stops the application from rendering translucent objects separately, which can be expensive for mobile devices because of fill rate limits~\cite{translucency}. Listing~\ref{list:framerate} shows such an example where developers enabled MobileMultiView to enable the path for Stereo Rendering on the mobile device's CPU. The fix also enabled Instance Stereo to help improve the performance of VR in UE.


{\fontsize{11}{11}
\begin{lstlisting}[columns=flexible,basicstyle=\small,caption=An Example of Instance Stereo and Mobile Multiview Fix \footnotesize{(microsoft/
MixedReality-UXTools-Unreal, 6e86e5a)}, frame=tblr, captionpos=b, label={list:framerate},escapechar=!]
...
!\hlp{-   vr.MobileMultiView=False}!
!\hlp{-   vr.InstancedStereo=False}!

!\hlg{+   vr.InstancedStereo=True}!
vr.MultiView=False
!\hlg{+   vr.MobileMultiView=True}!
...
\end{lstlisting}
}

\noindent\textbf{Rendering (3 smells)}
This performance smell category is specifically in linked to rendering settings and was kept separate from the object rendering parent category which covers CPP smells. The performance smells in this category affect the rendering of game objects and harm an application's performance. The first smell in this category is disabling \textbf{Occlusion Culling} when working with mobile devices. Having this setting enabled causes increased GPU usage, which leads to a decreased performance on these types of devices~\cite{desiredframerate}. In Listing~\ref{list:Culling} developer disabled occlusion queries to improve rendering performance.

{\fontsize{11}{11}
\begin{lstlisting}[columns=flexible,basicstyle=\small,caption=An Example of Occlusion Culling Fix \footnotesize{(microsoft
/
MixedReality-Unreal-Samples, b3441f4)}, frame=tblr, captionpos=b, label={list:Culling},escapechar=!]
...
r.DiscardUnusedQuality=False
!\hlp{r.AllowOcclusionQueries=True}!
!\hlg{r.AllowOcclusionQueries=False}!
r.MinScreenRadiusForLights=0.030000
...
\end{lstlisting}
}

The second rendering settings code smell is to enable \textbf{Forward Rendering} when developing VR applications. This allows for faster rendering passes which can increase framerates, an important part of VR performance~\cite{desiredframerate}. The last category \textbf{Ray Tracing} allows the creation of interactive experiences with lighting effects comparable with those of many offline renderers in real time~\cite{raytracing}. Listing~\ref{list:forward} shows a performance fix that enabled Forward Rendering settings for shading.

{\fontsize{11}{11}
\begin{lstlisting}[columns=flexible,basicstyle=\small,caption=An Example of Forward Rendering Fix \footnotesize{(1runeberg
/
RunebergVRPlugin, 71f2337)}, frame=tblr, captionpos=b, label={list:forward},escapechar=!]
...
[/Script/Engine.RendererSettings]
vr.InstancedStereo=True
!\hlg{+ r.ForwardShading=True}!

[/Script/Engine.PhysicsSettings]
DefaultGravityZ=-980.000000
...
\end{lstlisting}
}

\subsubsection{CPP Code Smells}
This category focuses on smells that can be found in the C++ source files of a project's directory and not the rendering project settings. There are two code smell categories associated with CPP files: \textbf{Motion Controller} transform updates and instanced \textbf{Static Mesh}.

\noindent\textbf{Motion Controller}
This category specifically focuses on \textbf{motion controllers} and the updates they cause as the user interacts with them. Low latency updates cause the render transforms to update twice per frame; this problem is prevalent when using motion controls as it can cause controller movements to appear jittery to the user~\cite{motioncontroller}. Disabling low latency updates on both motion controls removes the double updates and eliminates the performance issues that come with it. For example, in Listing~\ref{list:motioncontroller} the developer disables the motion controller to optimize performance.

{\fontsize{11}{11}
\begin{lstlisting}[columns=flexible,basicstyle=\small,caption=An Example of Forward Transformed  Update Fix\\ (\footnotesize{PitchAndYaw/ListenToMyVoice
, 2f4194e)}, frame=tblr, captionpos=b, label={list:motioncontroller},escapechar=!]
...
void AVRCharacter::BuildLeft() {
    _LeftHandComp = CreateDefaultSubobject<UMotionControllerComponent>(TEXT("\_LeftHandComp"));
!\hlg{+    \_LeftHandComp->bDisableLowLatencyUpdate = true;}!
    _LeftHandComp->Hand = EControllerHand::Left;
    _LeftHandComp->AttachToComponent(_VROriginComp, FAttachmentTransformRules::KeepRelativeTransform);
    _LeftHandComp->SetRelativeLocation(FVector(10.f, 0.f, 0.f));
...
\end{lstlisting}
}


\noindent\textbf{Static mesh}
Using static meshes in an XR project can be taxing on the GPU because each one needs to be rendered separately. The suggested fix for this code smell is to use an \textbf{instanced static mesh} instead. This allows identical meshes to be rendered in a single draw call, making them much more efficient than standard static meshes, especially when multiple meshes are needed. There are possible drawbacks to this solution that developers need to be aware of, such as the fact that they can not be individually culled and cannot be changed at runtime~\cite{staticmesh1}.
Textures should also be considered when developing for performance in XR applications. Using too many individual textures will cause an increase in draw calls and a decrease in performance. Limiting the use of individual textures can be accomplished by using smaller working sets, texture compression, or by using texture atlases~\cite{desiredframerate}. Example~\ref{list:staticmesh} shows such a fix where the developer loaded the mesh at the beginning of the play event of the actor and later re-used the mesh.

{\fontsize{11}{11}
\begin{lstlisting}[columns=flexible,basicstyle=\small,caption=An Example of Instanced Static Mesh Fix\footnotesize{Prastiwar/
UnrealArcheryShooter, 6b10e98)}, frame=tblr, captionpos=b, label={list:staticmesh},escapechar=!]
...
Super::BeginPlay();
...
!\hlg{+	// spawn instanced static mesh actor, better performance than individual actors}!
!\hlg{+			PaintMaterialInstance = world->SpawnActor<APaintMaterial>(PaintMaterial);}!
...
\end{lstlisting}
}

\subsection{RQ2: Information Source for Smell Detection}
\begin{figure*}[!htbp]
\centering
 \includegraphics[width=0.9\linewidth ]{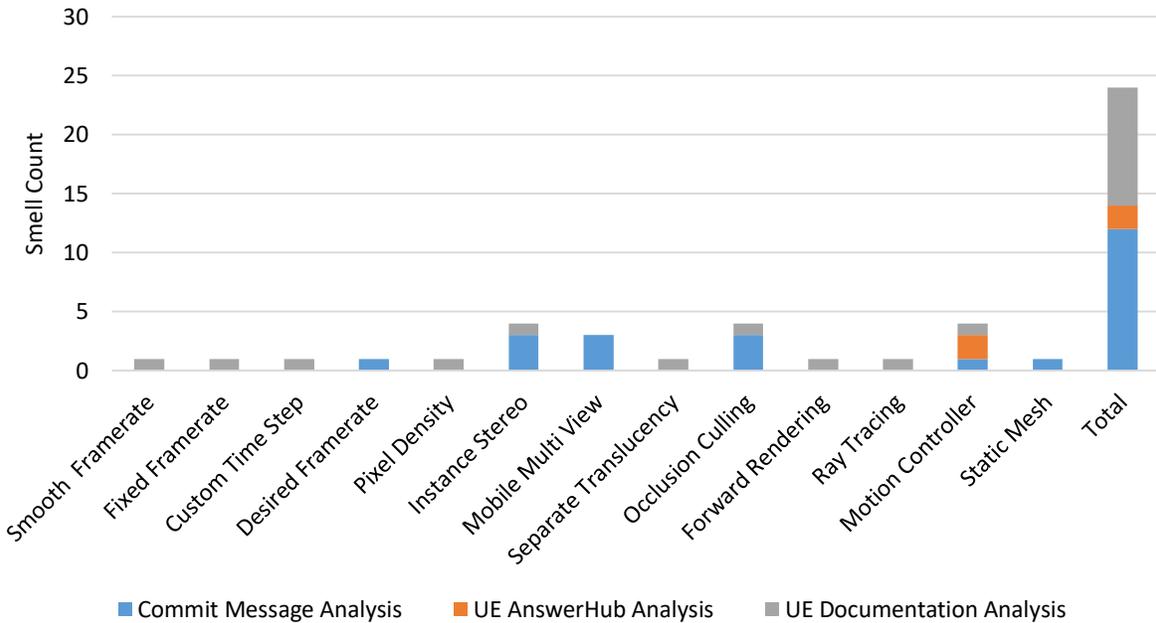}
 \caption{Information Source Distribution for Performance Smells}
 \label{fig:smellsource}
\end{figure*}

To identify UE XR performance smell, we analyzed three categories of information sources: 1) Code commits, 2) Issues from UE AnswerHub and 3) UE documentation. Figure~\ref{fig:smellsource} shows the performance smell source distribution. Based on our analysis, we identified six different categories of performance, including four related to settings smell and two related to CPP smell categories. In total, we identified 12 performance smells within these six categories. From UE AnswerHub, we were able to detect one category of performance smell, which is CPP Motion Controller. We identified two Motion Controller smells by analyzing issues. On the other hand, by analyzing UE Documentation, we identified nine categories of settings smell and one category of CPP performance smell. As UE documentation discussed each category of the smell once, from these ten categories of smell, we identified ten distinct performance smells. Even though UE documentation presented many of the smells for developers, three smells were never discussed in it. However, they were encountered by the developers as we discovered our through analysis of the performance commits. These are, Desired Framerate, Mobile Multiview and Static Mesh performances issues. At the same time, we observed one interesting point which is that in UE AnswerHub, developers discussed very little about performance issues. From our analysis, it's evident that developer commits and UE documentation are the most prevalent sources of information to know more about UE XR performance smells. 

\subsection{RQ3: Effectiveness of UEPerfAnalyzer  for  Performance-Smells Detection}

To evaluate the effectiveness of UEPerfAnalyzer's smell detection, we followed a two prong approach. First, a total of 100 smells were selected from the set of the automatically selected smells, with ten smells corresponding to each of the 8 INI file smells and 20 smells corresponding to the CPP smells. The CPP smells were composed of 19 Instanced Static Mesh Code and 1 Motion Controller Code Smell since only one instance was found of the latter. Then two co-authors manually evaluated these smells, marking them as TP ( True Positive) or FP ( False Positive). Agreement between them was at 98\% with a Cohen's Kappa of 0.78, and any differences were resolved via discussion. Furthermore, in order to find possible False Negatives, the co-authors also analyzed the INI and CPP files of the 'GabrielPaliari/pokeAR' project to see if there were any smells within this project that were not detected by \tool, but none were found. In total, the evaluation process produced 94 TP, 6 FP, 0 FN and 0 TN, resulting in a recall of 1, accuracy of 0.94 and F-1 score of 0.96.

By evaluating the automatic results using the manual results as a baseline, we found that the recall was 1, the accuracy was 0.94, and the F-1 score was 0.96. 

\subsection{RQ4: Distribution of performance smells Detected by UEPerfAnalyzer}
\begin{figure*}[!htbp]
\centering
 \includegraphics[width=0.8\linewidth,height=7cm ]{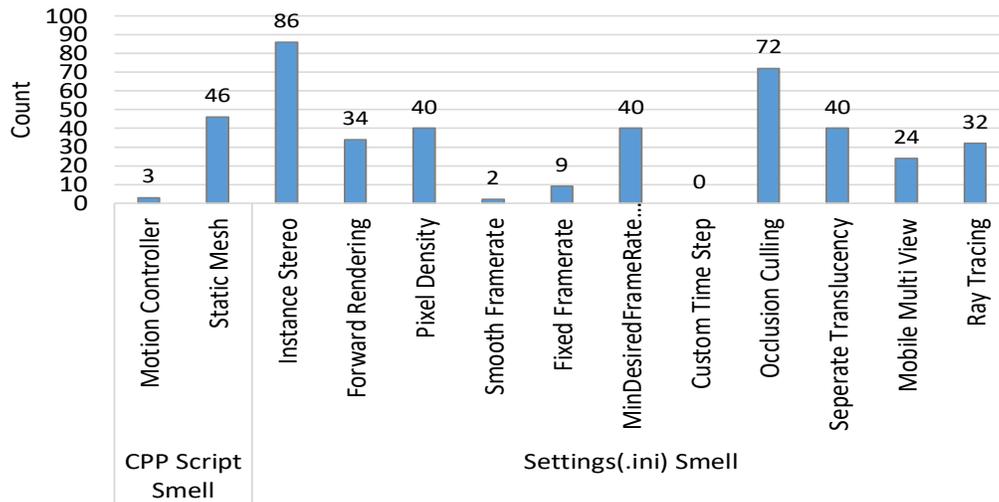}
 \caption{Detected Performance Smell Distribution}
 \label{fig:smelldistribution}
\end{figure*}

To understand the frequency of XR performance smells in practice, we analyzed 37 UE-based XR applications that we extracted while preparing the dataset and were not used for taxonomy generation.
During the analysis, UEPerfAnalyzer analyzed 489 C++ script files and 471 settings(.ini) files. Figure~\ref{fig:smelldistribution} shows the distribution of performance smells detected in evaluation projects. Based on the analysis result, we identified 49 C++ script-related smells that belonged to two categories of smells. Among these two categories of CPP script smells, Instanced Static Mesh Code Smell seems the most frequent one and accounts 93.88\% of C++ code smell. From settings files, UEPerfAnalyzer identified 379 performance smells. Among these settings-related smells, Instance Stereo and Occlusion Culling smells are the most prevalent smell types. Apart from that, Forward Shading, Pixel Density, MinDesiredFrameRate Framerate and Ray Tracing smells are also identified frequently. 

\begin{figure}[!htbp]
\centering
 \includegraphics[width=0.8\linewidth,height=6.5cm ]{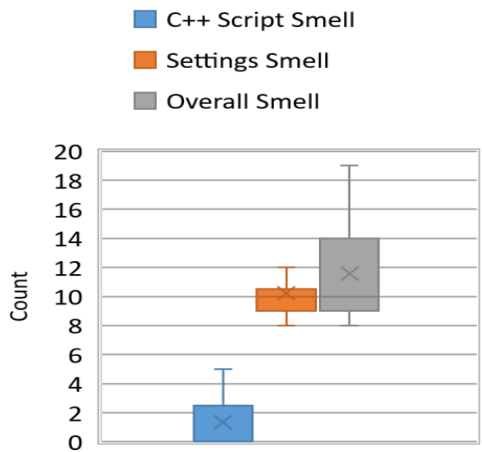}
 \caption{Project-wise smell distribution (omitting outlier)}
 \label{fig:projsmelldistribution}
\end{figure}

Figure~\ref{fig:projsmelldistribution} shows project-wise smell distribution of C++ script smell and settings smells. Based on the analysis, we identified that for each project the median number of CPP scripts is 1.32 and the median number of settings smell is 10.24. Overall, each of the projects contains more than 11 performance smells that show necessity to have performance bottleneck detection tools such as \tool to improve the performance of XR application. Among the analyzed projects, \texttt{sandisk/GabrielPaliari} project has 60 settings smell which is the highest among the analyzed projects. In the case of CPP script smells, \texttt{sandisk/BillyOMahony} has the highest number of performance smells with nine smells. 

%% file: Implications.tex
\section{Implications}
\label{sec:implications}
\subsection{For Extended Reality (XR) Developers}
\par{Our results point to many factors that Unreal Engine XR developers need to keep in mind during the development process. These factors have led us to the following concrete suggestions:
\begin{itemize}
    \item Unreal Engine project settings need to be finely tuned regarding object rendering. Disabling/enabling many of the settings (occlusion culling, forward rendering, distance culling, ray tracing, etc.) can help performance on specialized hardware needed for XR applications.
    \item Framerate settings are also of great importance depending on the XR application being developed. Disabling settings like smooth framerate, fixed framerate, and custom timesteps allow developers to control the framerate of their application directly. Most VR headsets are 90hz, hence setting the desired min/max framerates to 90+ is essential to maintaining good performance in these types of applications.
    \item Optimization through C++ source code is also important when it comes to XR application performance. To help balance performance issues that can occur from the unique hardware constraints that apply during XR development, the following settings recommendations are given: Disabling low latency on motion controllers, setting distance culling on objects, and using instance static meshes instead of standard static meshes. 
\end{itemize}
}
\subsection{For Software Engineering Researchers}
\par{Our research opens up potential research opportunities for software engineering researchers that can lead to improved XR development as well as better general Unreal Engine development.}  
\par{\textbf{Research on Unreal Engine project settings for general development.} Our findings point heavily in the direction of Unreal Engine project settings and the .ini files that contain them. Twelve of the fifteen code smells found in our taxonomy were related to project settings and can be resolved by updating the .ini files within a project's directory. Researching the project settings in a more general sense can lead to further findings on the importance of these files.}  
\par{\textbf{Further focus on the Virtual Reality (VR) aspects of the research.} A large portion of our findings were connected to the VR side of Extended Reality (XR) development. Performance considerations during the development process, including rendering and display settings as well as CPP optimization, are important because of the unique nature of VR headsets. The stereoscopic functionality of VR hardware relies on these optimizations as they need to render two versions of the virtual environment simultaneously. Moreover, researchers should extend further research on mining code commits and UE documentation to identify more performance issues XR applications
}

%% file: Discussion.tex
\section{Threats to Validity}
\label{sec:threatsvalidity}
Our analysis has some limitations that we would like to discuss:

\textbf{Internal validity:} The major threat to the internal validity of our study is the correctness of our manual inspection of performance commits, issues and documentation. To reduce this threat, three experts evaluated all the selected commits, issues and documentations separately and discussed them thrice to reach a consensus. Furthermore, the UEPerfAnalyzer tool developed for detecting performance smell might exhibit false results. However, to mitigate such a threat, we evaluated the tool on manually labeled data. The high accuracy result of UEPerfAnalyzer mitigates the threat and supports high confidence in performance smell detection results.

\textbf{External validity:} Our analysis is based on public repositories on UE-based XR applications collected from GitHub. The results might differ for privately developed XR applications and closed repositories, including commercial projects. Since we searched UE applications exhaustively and collected projects that are prevalent in terms of XR application standards, we expect similar performance smell behavior for privately developed XR applications. Apart from that, our analysis focused mainly on UE-based XR applications developed in C++. As UE is one of the most popular frameworks for XR application development, we expect that the performance issues we observed will be beneficial for XR developers and researchers in optimizing performance for XR applications.

%% file: Relatedworks.tex
\section{Related Works}
\label{sec:relatedworks}
\subsection{Studies on Performance Bug Analysis}
Since Performance is one of the top most non-functional requirements for software application~\cite{Kim2016FSE,Han2018ASE}, several studies analyzed performance bottlenecks and optimization techniques for different software variants. Prior work by Liu et al.~\cite{Liu2014ICSE} performed empirical analysis on Android applications. Their work also proposed a static analysis tool PerfChecker, to detect identified performance bug patterns. The work Zaman et al.~\cite{Zaman2012MSR} performed qualitatively studies a random sample of 400 performance and non-performance bug reports of Mozilla Firefox and Google Chrome. The study pointed out that performance issues are difficult to reproduce and also require more discussion to fix the performance issues. Nistor et al.~\cite{Nistor2013MSR} also performed a similar study on performance and non-performance bugs from three popular codebases: Eclipse JDT, Eclipse SWT, and Mozilla. The work summarized that fixing performance bugs is more challenging than non-performance bugs. Moreover, fixing performance bugs may introduce new functional bugs than fixing non-performance bugs, which implies the complexity of fixing performance bugs. Selakovic and Pradel et al.~\cite{Selakovic2016ICSE} performed empirical studies on 98 fixed performance issues from 16 popular client-side and server-side JavaScript projects. The work identified eight root causes of performance issues with inefficient usage of APIs being the most prevalent root cause of performance bugs. A recent study by Chen et al.~\cite{Chen2019ASE} on database-centric Web Applications identified 17 performance
anti-patterns, from industrial web applications written in Laravel, which is the most popular web framework in PHP. 
To assist developers in writing test cases for performance bugs, Han et el.~\cite{Han2018ASE} proposed an approach that can analyze how performance bugs are related to the combinations of input parameters. Based on their understanding, they proposed an approach named PerfLearner that can generate test frames for the bug reports. Mostafa et al.~\cite{Mostafa2017ISSTA} proposed PerfRanker for prioritizing performance regression test cases in collection-intensive software. Recently, Garg et al.~\cite{Garg2021} proposed PerfLens:  a data-driven approach to software performance improvement in C\#. PerfLens can suggest performance improvements with 90\% accuracy with available profiler data and 55\% accuracy with the availability of source code only. Several other researchers~\cite{Olivo2015pldi,Killian2010,Nistor2013ToddlerDP,Nistor2015CARAMELDA} developed tools and techniques that can detect performance bugs in software applications. Prior studies mostly focus on generic software performance issues, which in many cases can not be directly applicable to XR applications. In our research, we focused on performance optimizations of XR applications and developed an analyzer that can detect performance smell in UE-based XR applications.

\subsection{Studies on XR and Game Development}
With the growing popularity of XR applications, including VR, AR and MR and Game applications, the research community recently focused more on studying XR and Game applications. Truelove et al.~\cite{Truelove2021} performed empirical analysis on 12,122 bug fixes from 723 updates of 30 popular games on the Steam platform to generate bug taxonomy for Game applications. The work also included Game developer surveys to identify the different aspects that can be associated with different bug types. Their analysis identified that Information bugs appear most frequently in updates, while Crash bugs recur the most frequently and are often treated as more severe than other bug types. Wu et al.~\cite{Wu2020ICMSE} proposed a machine learning-based regression testing technique for detecting regression bugs for Games. Recently, Politowski et al.~\cite{Politowski2021ASO} performed a survey to understand existing testing processes and how they could automate them. Their analysis discovered that game developers rely almost exclusively on manual play-testing and are highly dependent on a tester's knowledge. Analyzing the commit history of open-source Unity virtual reality projects to identify the types of optimizations developers perform revealed that the top categories included graphics simplification, rendering optimization, and language feature/api optimizations~\cite{NusratICSE2021}. By widening the project focus group from virtual reality applications to encompass all open source game projects hosted on GitHub, 1249 anti-patterns were identified and categorized into ten themes while analyzing commits, issues, and pull requests from 100 repositories~\cite{AgrahariACM2022}. Our work is similar in nature, utilizing open-source projects' commit history to identify potential code smells. We took a different approach focused on the identification of code smells  within Unreal Engine open source XR Projects. Unity and Unreal Engine are both widely used game engines, however, work related to our research tends to be focused on Unity projects. A proposed project, UnityLinter, is a tool developed to perform static analysis to identify code smells in Unity-based projects~\cite{BorrelliACM2020}. The code smells that it is capable of detecting were identified by analyzing 100 open source repositories~\cite{BorrelliACM2020}. Our detection system allowed us to obtain a list of potential code smells through the use of static analysis; however, we focus on performing static analysis on UE XR-specific projects, which are  vastly different architecture than projects developed in Unity.

%% file: Conclusion.tex
\section{Conclusion}
\label{sec:conclusion}
In this study, we performed an empirical study on 279 performance commits collected from 7 UE-based XR projects, 78 performance-related issues collected from UE AnswerHub and three sources of UE documentation. With the manual analysis, we developed a performance bug taxonomy for UE XR applications with 14 distinct types of performance bug types involving CPP script and settings configurations. Our analysis of finding performance smell sources indicates that developer commits and UE documentation are the most prevalent source of information to know more about performance smells and their fixes. Moreover, to assist the developer, we developed a static analyzer \tool that can detect UE performance code smells with high accuracy. For the 37 evaluation projects, \tool detected  49 CPP script-related performance smells and 379 settings-related performance smells. To the best of our knowledge, this is the first research work on UE-based XR performance bug analysis and also developed tool support for developers to detect those performance smells. We plan to extend our dataset to perform to validate our study further.  At the same time, we plan to extend our analysis to analyze performance issues in UE asset and prefab files and develop tool support to detect performance issues related to UE asset and prefab files. Moreover, we plan to develop tool support to assist developers with possible performance fix recommendations.